\documentclass[3p, 11pt]{elsarticle}
\usepackage{graphicx,amscd,amsmath,amssymb,verbatim}
\usepackage{amsfonts,epsfig}
\usepackage{mathptmx}
\usepackage{setspace}
\usepackage{epsfig}
\usepackage{url}
\usepackage{multirow}
\usepackage{color}
\usepackage{subfigure}

\begin{document}

\journal{Elsevier}

\begin{frontmatter}

\title{A new look on the stabilization of inverted pendulum with parametric
excitation and large random frequencies: analytical and numerical approaches}

\author{Roberto da Silva$^1$, Sandra D. Prado$^1$, Henrique A. Fernandes$^2$} 

\address{$^1$Instituto de F{\'{\i}}sica, Universidade Federal do Rio Grande do Sul,
Av. Bento Gon{\c{c}}alves, 9500 - CEP 91501-970, Porto Alegre, Rio Grande do
Sul, Brazil\\
$^2$Coordena{\c{c}}{\~a}o de F{\'i}sica, Universidade Federal de Goi{\'a}s,
Regional Jata{\'i}, BR 364, km 192, 3800 - CEP 75801-615, Jata{\'i}, Goi{\'a}%
s, Brazil\\
{\normalsize{E-mail:rdasilva@if.ufrgs.br}}}

\begin{abstract}

In this paper we explore the stability of an inverted pendulum with
generalized parametric excitation described by a superposition of $N$ sines
with different frequencies and phases. We show that when the amplitude is
scaled with the frequency we obtain the stabilization of the real inverted
pendulum, i.e., with values of $g$ according to planet Earth ($g\approx 9.8$
m/s$^{2}$) for high frequencies. By randomly sorting the
frequencies, we obtain a critical amplitude in light of perturbative theory
in classical mechanics which is numerically tested by exploring its validity
regime in many alternatives. We also analyse the effects when different values 
of $N$ as well as the pendulum size $l$ are taken into account.

\end{abstract}


\end{frontmatter}

\tableofcontents

\setlength{\baselineskip}{0.7cm}

\section{Introduction}

\label{Section:Introduction}

The inverted pendulum and its stability are subjects widely explored in
Physics, Engineering, Biology \cite{Ibrahim2006}, and many other areas due
to its technological importance. An inverted pendulum is unstable unless
some kind of excitation/vibration is applied to its suspension point (its
basis). Kaptiza \cite{Kaptiza1951,Kaptiza1965} observed that an inverted
pendulum should be stabilized by rapidly oscillating its basis. The limit of
stability considering a periodic function at the basis of pendulum has been
studied by many authors (see for example \cite{Landau,Butkov2001,Erdos1996}%
). In another context, chaos and bifurcations have been studied for a
sinusoidal excitation where both excitation frequencies and amplitudes were
varied \cite{Kim1998}. However, this excitation can be more general openning
a long way to explore the stochastic aspects in the stabilization.

By denoting $z(t)$ as a vertical excitation, the Lagrangian of a pendulum
with mass $m$ can be written as 
\begin{equation}
\begin{array}{lll}
\mathcal{L}(\theta ,\dot{\theta},z,\dot{z}) & = & \frac{1}{2}ml^{2}\left( 
\dot{\theta}^{2}+\frac{1}{l^{2}}\dot{z}^{2}(t)\right) -ml\dot{z}(t)\dot{%
\theta}\sin \theta \\ 
&  & -mgz(t)-mgl\cos \theta%
\end{array}
\label{Eq:Lagrangian}
\end{equation}%
where the axis $z$ is oriented up, $\vec{a}=-g\hat{z}$ is the gravitational
acceleration, and $l$ is the pendulum length, which in turn, leads to the
following equation of motion 
\begin{equation}
\frac{d^{2}\theta }{dt^{2}}=\frac{g}{l}\left( 1+\frac{1}{g}\ddot{z}\right)
\sin \theta .  \label{Eq:geral}
\end{equation}

Assuming $\overset{..}{z}$ as a generic time-dependent function and taking
into account the limit of small oscillations $\sin \theta \approx \theta $,
one has 
\begin{equation}
\ddot{\theta}=\left( \omega _{0}^{2}+\frac{1}{l}\ddot{z}\right) \theta ,
\label{Eq:Pendulum}
\end{equation}%
where $\omega _{0}^{2}=g/l$.

An interesting choice is to consider a parametric excitation 
\begin{equation}
z(t)=\sum_{i=1}^{N}A_{i}\sin (\omega _{i}t+\varphi _{i}),
\label{Eq:parametric}
\end{equation}%
where $A_{i}$, $\omega _{i}$, and $\varphi _{i}$ are arbitrary amplitudes,
frequencies and phases, respectively, and $i=1,...,N$. A detailed study of
this kind of excitation in an inverted pendulum can be found, for example,
in Refs. \cite{Ibrahim2006,Rdasilva2016}. In a very different context,
Dettmann \textit{et al.} \cite{Prado2004} had obtained an equation for the
distance between two photons that propagate in a universe of negative
curvature. This distance can be written as 
\begin{equation}
\ddot{\theta}=\left( 1-f(t)\right) \theta .  \label{Eq:Cosmological}
\end{equation}%
At a first glance, this result seems to be a particular case of Eq. (\ref%
{Eq:Pendulum}), where $\omega _{0}^{2}=1$ and $f(t)$ is a stochastic forcing
function that, in their particular case, takes into account the perturbation
in the curvature due to mass distribution $\omega _{0}^{2}=1$. They studied
the stochastic stabilization of Eq. (\ref{Eq:Cosmological}) by considering 
\begin{equation}
f(t)=f_{D}(t)=A\sum_{i=1}^{N}\sin (\omega _{i}t+\varphi _{i}),
\label{Eq.f_para_Dettman}
\end{equation}%
where $A$ is a control parameter and $\left\{ \omega _{i},\varphi
_{i}\right\} _{i=1}^{N}$ are chosen independently at random according to a
uniform distribution defined on supports: $[\omega _{\min },\omega _{\max }]$
and $[\phi _{\min },\phi _{\max }]$ respectively.

It is important to emphasize that by simply making $\omega _{0}^{2}=1$ in
Eq. (\ref{Eq:Pendulum}) with $A_{i}=A$ in Eq. (\ref{Eq:parametric}), we do
not recover Eq. (\ref{Eq:Cosmological}), since the usual $f(t)$ considered
for a regular pendulum is 
\begin{equation}
f(t)=f_{S}(t)=-\frac{1}{l}\overset{..}{z}(t)=\frac{1}{l}\sum_{i=1}^{N}A_{i}%
\omega _{i}^{2}\sin (\omega _{i}t+\varphi _{i}).
\label{Eq:Silva_Debora_Sandra}
\end{equation}

At this point two technical problems occur. Firstly, $\omega _{0}^{2}=1$
means the specific case of a huge pendulum ($l\approx 9.8$ m). Secondly, we
should incorporate the gravity $g$ in $A$. However, the term $\omega
_{i}^{2} $ does not exist in the original problem considered by Dettman 
\textit{et al.} \cite{Prado2004}.

The main contributions of this paper is related to the stabilization of the
inverted pendulum or similar system. Here, we answer the following two
questions and compare both situations:

\begin{enumerate}
\item Is it possible to stabilize an inverted pendulum in a general
situation, i.e., by considering the Eq. (\ref{Eq:Pendulum}) with the
parametric excitation $z(t)=A\sum_{i=1}^{N}\sin (\omega _{i}t+\varphi _{i})$
with random frequencies uniformily distributed in $[\omega _{\min },\ \omega
_{\max }]$? If yes, what should be the parameters $\omega _{0}$ and $A$?

\item Is it possible to obtain a more general stabilization criteria by
considering any values of parameters $\omega _{\min },\ \omega _{\max }$ and 
$A$? This question arises because Dettman \textit{et al.} showed that the
problem for a particular cosmological application (Eqs. (\ref%
{Eq:Cosmological}) and (\ref{Eq.f_para_Dettman})) can be \textquotedblleft
stochastically\textquotedblright\ stabilized when considering a specific
choice of $\omega _{\min },\ \omega _{\max }$ and $A$.
\end{enumerate}

Throughout this work we will present the answer of these questions. However,
we would like to point out that the answer is yes, we are able to stabilize
the inverted pendulum by considering more general criterias.

Our manuscript is organized as follows: In the next section, we present the
perturbative calculations in detail and show a general solution for the
problem. For this purpose, we consider the more general equation 
\begin{equation}
\ddot{\theta}=\left( \omega _{0}^{2}-f(t)\right) \theta   \label{Eq:General}
\end{equation}%
with a more general function 
\begin{equation*}
f(t)=\sum_{i=1}^{N}A_{i}^{\ast }\sin (\omega _{i}t+\varphi _{i})
\end{equation*}%
where $\omega _{0}^{2}=1$ and $A_{i}^{\ast }=A$ correspond to the
cosmological problem (Problem I) and $\omega _{0}^{2}=\frac{g}{l}$ and $%
A_{i}^{\ast }=\frac{\omega _{i}^{2}}{l}A_{i}$, $i=1,2,...,N$ correspond to
the general regular inverted pendulum (Problem II). In Sec. \ref%
{Section:Results} we show our results and the conclusions are presented in
Sec. \ref{Section:Results}.

\section{Perturbative methods}

\label{Section:Perturbative_methods}

In this section we describe in detail how to obtain a general stability
condition to the Eq. (\ref{Eq:General}). An interesting Ansatz to start with
is: 
\begin{equation}
\theta (t)=\phi _{slow}(t)+\omega ^{-\alpha }\phi _{fast}(t)
\label{Eq:General_solution}
\end{equation}%
where the motion in decomposed in two parts: one \textit{slow} and another 
\textit{fast}. The fast part corresponds to an additional noise to the main
motion (the slow one). The \textit{ad hoc} parameter $\omega ^{-\alpha }$
controls the contribution of the fast component, and $\alpha $ is a positive
number which characterizes this term. The quantity $\omega =2\pi /T$\ is an
average over the differents $\{\omega _{i}\}_{i=1}^{N}$.

By substituting Eq. (\ref{Eq:General_solution}) into Eq. (\ref{Eq:General})
we obtain 
\begin{equation}
\begin{array}{lll}
\overset{\cdot \cdot }{\phi }_{slow}(t)+\omega ^{-\alpha }\overset{\cdot
\cdot }{\phi }_{fast}(t) & = & \omega _{0}^{2}\phi _{slow}(t)-f(t)\phi
_{slow}(t) \\ 
&  & +\omega ^{-\alpha }\omega _{0}^{2}\phi _{fast}(t) \\ 
&  & -\omega ^{-\alpha }f(t)\phi _{fast}(t)%
\end{array}
\label{Eq:original}
\end{equation}

Now it is crucial to consider the nature of motion to distinguish the
important terms in Eq. (\ref{Eq:original}). The only candidates associated
with the perturbative effects on the right side of this equation are $%
-f(t)\phi _{slow}(t)$, $\omega ^{-\alpha }\omega _{0}^{2}\phi _{fast}(t)$,
and $\omega ^{-\alpha }f(t)\phi _{fast}(t)$. Therefore, given that the terms 
$\omega ^{-\alpha }\omega _{0}^{2}\phi _{fast}(t)$\ and $\omega ^{-\alpha
}f(t)\phi _{fast}(t)$\ are small when compared with $-f(t)\phi _{slow}(t)$\
we have%
\begin{equation}
\omega ^{-\alpha }\overset{\cdot \cdot }{\phi }_{fast}(t)\approx -f(t)\phi
_{slow}(t).  \label{Eq. Original2}
\end{equation}

Let us define the time average as 
\begin{equation}
\left\langle g\right\rangle (t)=\mathbf{\ }\int_{-\infty }^{\infty }g(\tau
)\Phi _{T}(\tau -t)d\tau  \label{Eq:time_average}
\end{equation}%
where,

\begin{equation}
\Phi _{T}(\tau -t)=\left\{ 
\begin{array}{ccc}
1/T &  & \text{if }\left\vert \tau -t\right\vert <T/2 \\ 
&  &  \\ 
0 &  & \text{otherwise}%
\end{array}%
\right.  \label{Eq:Sequence}
\end{equation}%
where $T$ is small and by hyphotesis/construction $\left\langle \phi
_{fast}\right\rangle =0$. Here, $\lim_{T\rightarrow 0}\Phi _{T}(\tau
-t)=\delta (\tau -t)$, which corresponds to the limit $\omega \rightarrow
\infty $.

In this case one has

\begin{equation}
\left\langle \phi _{slow}(t)\right\rangle \approx \int_{-\infty }^{\infty
}\phi _{slow}(\tau )\delta (\tau -t)d\tau =\phi _{slow}(t)
\label{Eq:recaindo_na_delta}
\end{equation}%
and similarly 
\begin{equation}
\left\langle \overset{\cdot \cdot }{\phi }_{slow}(t)\right\rangle \approx
\int_{-\infty }^{\infty }\overset{\cdot \cdot }{\phi }_{slow}(\tau )\delta
(\tau -t)dt\tau =\ \overset{\cdot \cdot }{\phi }_{slow}(t).
\end{equation}

Now, with the choice of a fast oscillatory function $f(t)$\ such that $%
\left\langle f(t)\right\rangle =0$\ for some time interval $t$, and $%
\left\langle \phi _{slow}(t)\right\rangle \approx $\ $\phi _{slow}(t)$\ in
this time range, then it is easy to show that $\left\langle f(t)\phi
_{slow}(t)\right\rangle \approx \phi _{slow}(t)\left\langle
f(t)\right\rangle \approx 0$. After all these constraints, and taking the
time average according to Eq. (\ref{Eq:original}) the dynamics of the slow
component can be written as 
\begin{equation}
\overset{\cdot \cdot }{\phi }_{slow}(t)=\omega _{0}^{2}\phi
_{slow}(t)-\omega ^{-\alpha }\left\langle f(t)\phi _{fast}(t)\right\rangle .
\label{Eq:averaged}
\end{equation}

By integrating Eq. (\ref{Eq. Original2}) and having in mind that $\phi
_{slow}(t)$ has a slow variation, one has:

\begin{equation}
\omega ^{-\alpha }\overset{\cdot }{\phi }_{fast}(t)\approx -\phi
_{slow}(t)\int_{0}^{t}f(s)ds  \label{Eq:Auxiliar}
\end{equation}%
considering that $\overset{\cdot }{\phi }_{fast}(0)=0$. In the same way, by
integrating again yields 
\begin{equation}
\omega ^{-\alpha }\phi _{fast}(t)\approx -\phi _{slow}(t)x(t),
\label{Eq:intermediate}
\end{equation}%
where $x(t)=\int_{0}^{t}v(s)ds$ with $v(t)=\int_{0}^{t}f(s)ds$, and $\phi
_{fast}(0)=0$\ by hypothesis.

Multiplying the Eq. (\ref{Eq:intermediate}) by $f(t)$ and taking the average
again, one has: $\omega ^{-\alpha }\left\langle f(t)\phi
_{fast}(t)\right\rangle \approx -\phi _{slow}(t)\left\langle
f(t)x(t)\right\rangle $. Let us now calculate $\left\langle
f(t)x(t)\right\rangle $. According to Eq. (\ref{Eq:time_average}), this
average is given by 
\begin{equation*}
\left\langle f(t)x(t)\right\rangle =\frac{1}{T}\int_{0}^{T}\frac{dv}{dt}%
x(t)dt.
\end{equation*}

Integrating the above equation by parts one has 
\begin{equation*}
\begin{array}{lll}
\left\langle f(t)x(t)\right\rangle & = & \frac{1}{T}\int_{0}^{T}\frac{dv(t)}{%
dt}x(t)dt \\ 
&  &  \\ 
& = & \frac{1}{T}\left[ v(T)x(T)-v(0)x(0)\right] -\frac{1}{T}\int_{0}^{T}v(t)%
\frac{dx}{dt}(t)dt \\ 
&  &  \\ 
& = & -\frac{1}{T}\int_{0}^{T}v^{2}(t)dt=-\left\langle v^{2}(t)\right\rangle%
\end{array}%
\end{equation*}%
providing $\omega ^{-\alpha }\left\langle f(t)\phi _{fast}(t)\right\rangle
\approx -\phi _{slow}(t)\left\langle f(t)x(t)\right\rangle =\phi
_{slow}(t)\left\langle v^{2}(t)\right\rangle $. The Eq. (\ref{Eq:averaged})
can then be written as

\begin{equation}
\overset{\cdot \cdot }{\phi }_{slow}(t)=\left[ \omega _{0}^{2}-\left\langle
v^{2}(t)\right\rangle \right] \phi _{slow}(t).  \label{Eq:final}
\end{equation}

\section{Results}

\label{Section:Results}

Let us start by considering the problem of a real inverted pendulum, the
problem previously called problem II. So, $f(t)=\frac{1}{l}%
\sum_{i=1}^{N}A_{i}\omega _{i}^{2}\sin (\omega_{i}t+\varphi _{i})$ and $%
\omega _{0}^{2}=g/l$. In addition, it is convenient to consider $A_{i}=l%
\frac{A}{\omega _{i}^{2}}$ whereas we are looking for a general and broad
treatment for the two problems raised in this work, simultaneously.

This means that if the frequencies are chosen at random according to a
uniform probability distribution, the amplitudes must be scaled by dividing
them by the square of those frequencies.

Here it is also important to observe that $\left\langle
v^{2}(t)\right\rangle $ in Eq.(\ref{Eq:final}) is a time-dependent function,
since 
\begin{equation*}
v(t)=-A\sum_{i=1}^{N}\frac{1}{\omega _{i}}\sin (\omega _{i}t+\varphi _{i})
\end{equation*}%
The square of this equation yields $v(t)^{2}=A^{2}\sum_{i=1}^{N}\frac{1}{%
\omega _{i}^{2}}\sin ^{2}(\omega _{i}t+\varphi _{i})+A^{2}\sum_{i\neq
j=1}^{N}\frac{1}{\omega _{i}\omega _{j}}\sin (\omega _{i}t+\varphi _{i})\sin
(\omega _{j}t+\varphi _{i})$.

However, we can make an \textit{ad hoc} consideration, which is a kind of
\textquotedblleft mean-field approximation\textquotedblright, by changing $%
\left\langle v^{2}(t)\right\rangle $ by a simple time average. This means to
consider $\left\langle v^{2}(t)\right\rangle $ as a constant $\overline{v^{2}%
}$, i.e., 
\begin{eqnarray*}
\overline{v^{2}} &=&A^{2}\sum_{i=1}^{N}\frac{1}{\omega _{i}^{2}}\overline{%
\sin ^{2}(\omega _{i}t+\varphi _{i})} \\
&&+A^{2}\sum_{i\neq j=1}^{N}\frac{1}{\omega _{i}\omega _{j}}\overline{\sin
(\omega _{i}t+\varphi _{i})\sin (\omega _{j}t+\varphi _{i})}
\end{eqnarray*}%
which leads to%
\begin{equation*}
\begin{array}{lll}
\overline{v^{2}} & = & \frac{A^{2}}{2}\sum_{i=1}^{N}\omega _{i}^{-2} \\ 
&  &  \\ 
& \approx & \frac{N}{2(\omega _{\max }-\omega _{\min })}A^{2}\int_{\omega
_{\min }}^{\omega _{\max }}\omega ^{-2}d\omega \\ 
&  &  \\ 
& = & \frac{N}{2}\frac{A^{2}}{\omega _{\max }\omega _{\min }},%
\end{array}%
\end{equation*}
which, in turn, leads to the stability condition: 
\begin{equation*}
\omega _{0}^{2}<\frac{N}{2}A^{2}\frac{1}{\omega _{\max }\omega _{\min }}.
\end{equation*}

More simply, 
\begin{equation}
A\geq A_{c}(\omega _{\min },\omega _{\max })=\sqrt{\frac{2g}{Nl}\omega
_{\max }\omega _{\min }}.  \label{Eq:Critical_A_c}
\end{equation}

In order to verify our approach, we take into account frequencies chosen at
random in the interval $[\omega _{\min },\omega _{\max }]$ and look into the
stabilization of the inverted pendulum. Considering these frequencies, our
perturbative analysis provides a lower bond for the amplitude $A$ which is
tested through numerical simulations.

In our simulations, we numerically integrate the Eq. (\ref{Eq:General})
using the fourth-order Runge-Kutta method \cite{Press1992} with $%
f(t)=\sum_{i=1}^{N}A\sin (\omega _{i}t+\varphi _{i})$ starting from $\theta
_{0}=1^{o}\approx 0.018$ rad and $\dot{\theta}_{0}=0$.

First, we fix the amplitude $A$ and randomly sort $N$ frequencies $\omega
_{i}$, $i=1,...,N$ with uniform distribution in the interval $[\omega _{\min
},\omega _{\max }]$. In addition, we also sort $N$ random phases $\varphi
_{i}$, $i=1,...,N$ in the interval $[-\pi ,\pi ]$. Then, we repeat the
procedure $N_{run}$ times we test how many times the stability is broken
considering a maximum number of iterations $\tau _{\max }=10^{5}$.

Finally, we calculate the following quantities:

\begin{enumerate}
\item \textbf{survival probability}: It is denoted by $p=\frac{n}{N_{run}}$,
where $n\leq N_{run}$ is the number of times such that the time evolution
reaches the maximal number of iterations $\tau _{\max }$.

\item \textbf{survival time}: It is the time $\tau \leq \tau _{\max }$, and
describes the first time that the condition $\cos \theta <0$ is satisfied,
i.e., the pendulum loses its stability (which is an appropriated and robust
stability condition as discussed and numerically verified in Ref. \cite%
{Rdasilva2016}).
\end{enumerate}

\begin{figure*}[th]
\begin{center}
\includegraphics[width=0.5%
\columnwidth]{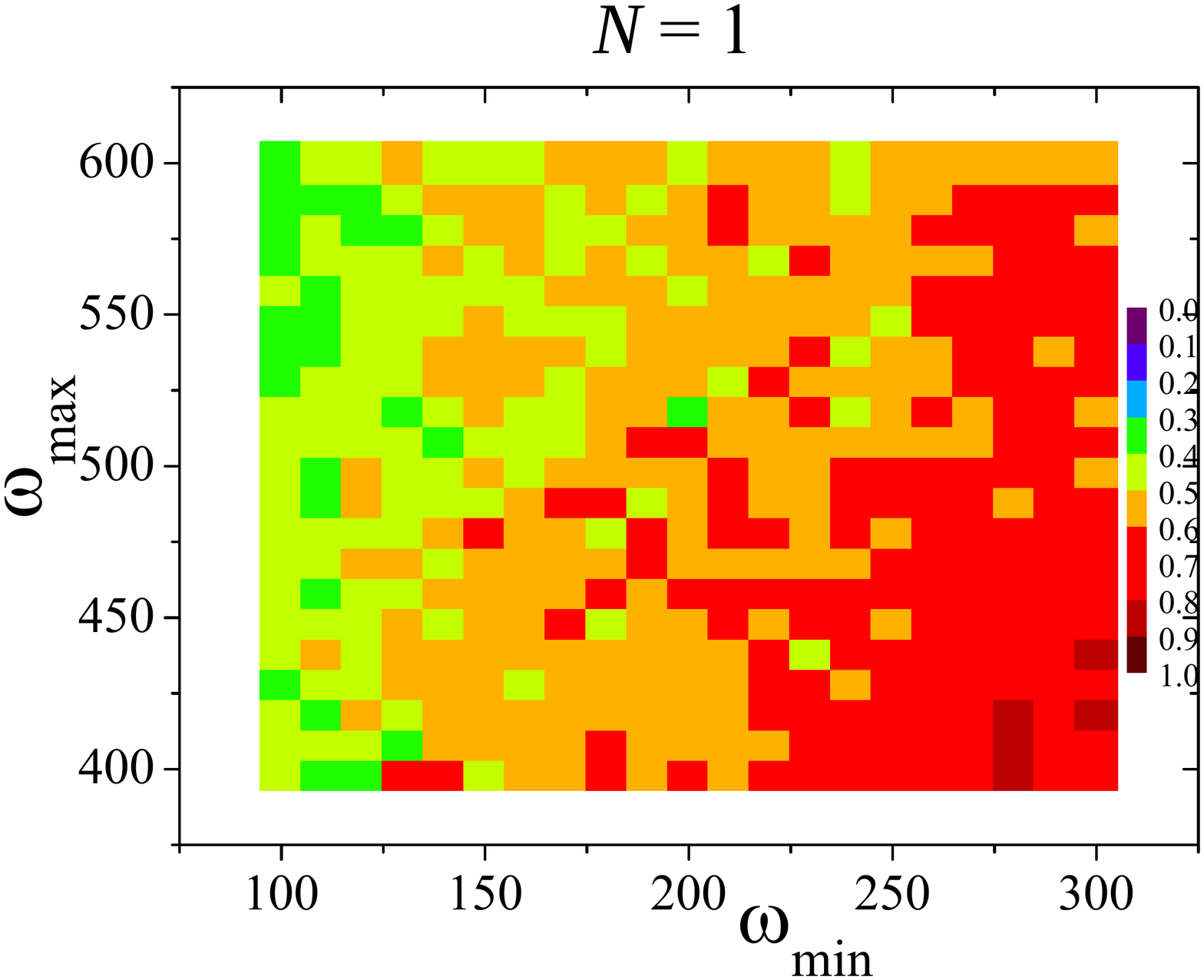}%
\includegraphics[width=0.5%
\columnwidth]{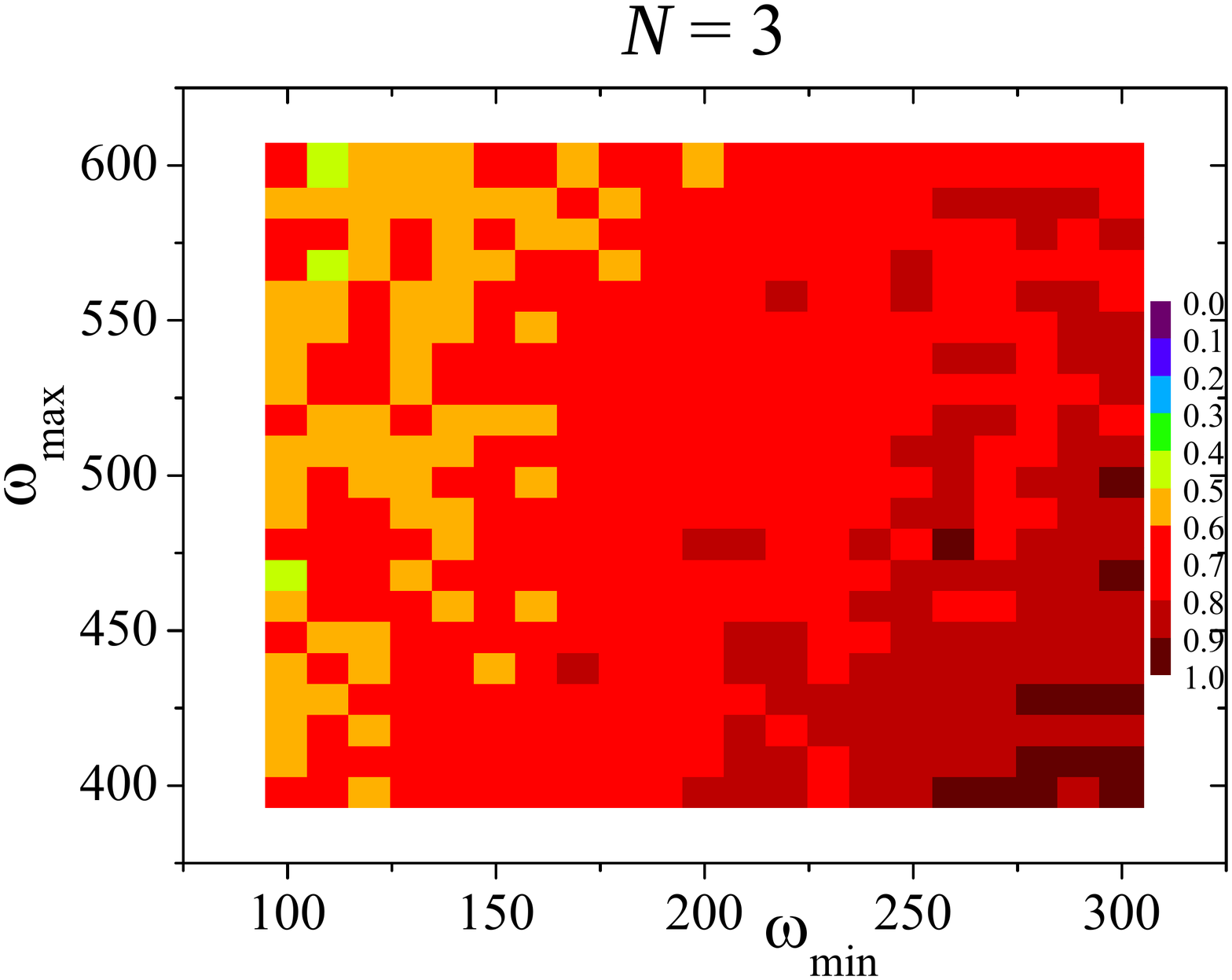} %
\includegraphics[width=0.5%
\columnwidth]{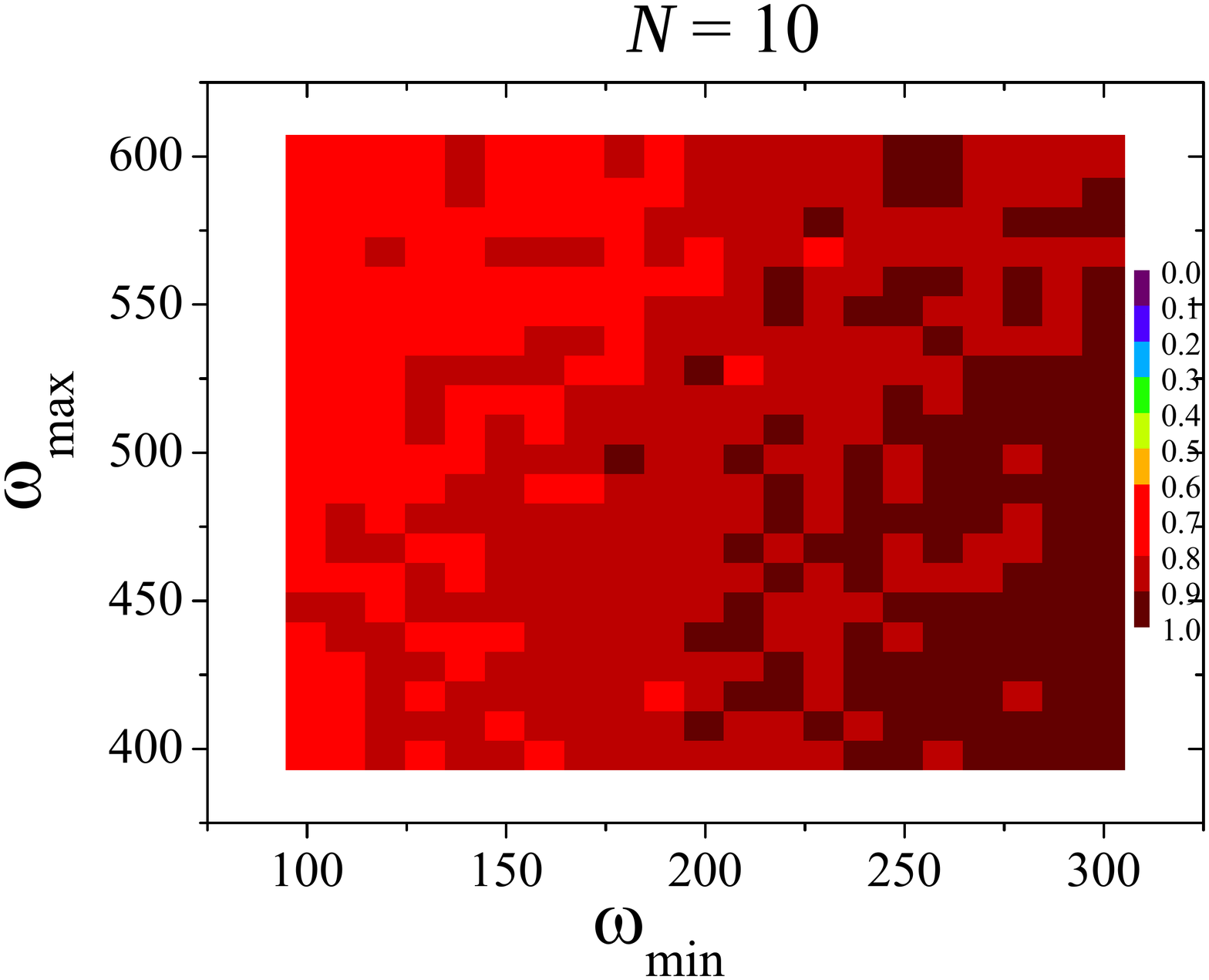}%
\includegraphics[width=0.5%
\columnwidth]{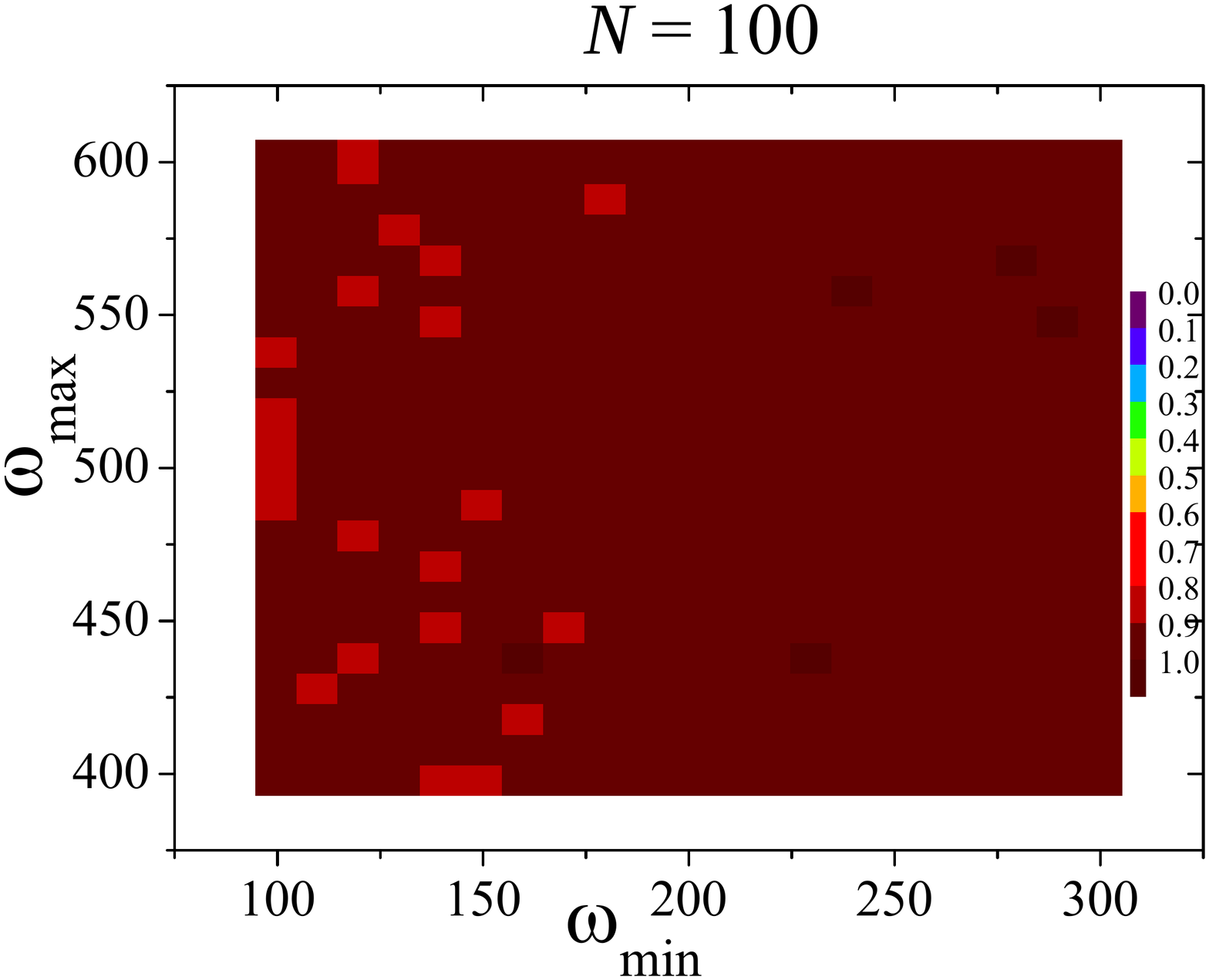}
\end{center}
\caption{Stability diagrams for $l\protect\overset{N}{=}g=9.8$ m/s$^{2}$
obtained by numerical integration. For each pair ($\protect\omega _{\min },%
\protect\omega _{\max }$) we draw the survival probability $p$ according to
rainbow collor scale (very cold $p\rightarrow 0$ and very hot $p\rightarrow
1 $). }
\label{Fig:Effects_number_cosine2}
\end{figure*}

In order to perform an exhaustive numerical analysis for each pair ($\omega
_{\min },\omega _{\max }$), we build stability diagrams considering the
number of sines being added in $f(t)$. Since the results from the
perturbative analysis must get better as $N$ increases, it is interesting to
study the influence of $N$ on the inverted pendulum stability. We consider
the system exactly in $A=A_{c}(\omega _{\min },\omega _{\max })$ (according
to Eq. (\ref{Eq:Critical_A_c})) which is the necessary minimum amplitude
according to the perturbative theory previously developed. It is important
to notice that our simulations use the critical minimal value $A_{c}$ for
the amplitude in order to test the robustness of our theoretical result
given by Eq. (\ref{Eq:Critical_A_c}), whereas larger amplitudes should bring
more stability. The idea of fixing $A=A_{c}(\omega _{\min },\omega _{\max })$
is interesting because it saves one dimension in the plots that are shown
below.

\begin{figure}[th]
\begin{center}
\includegraphics[width=1.0%
\columnwidth]{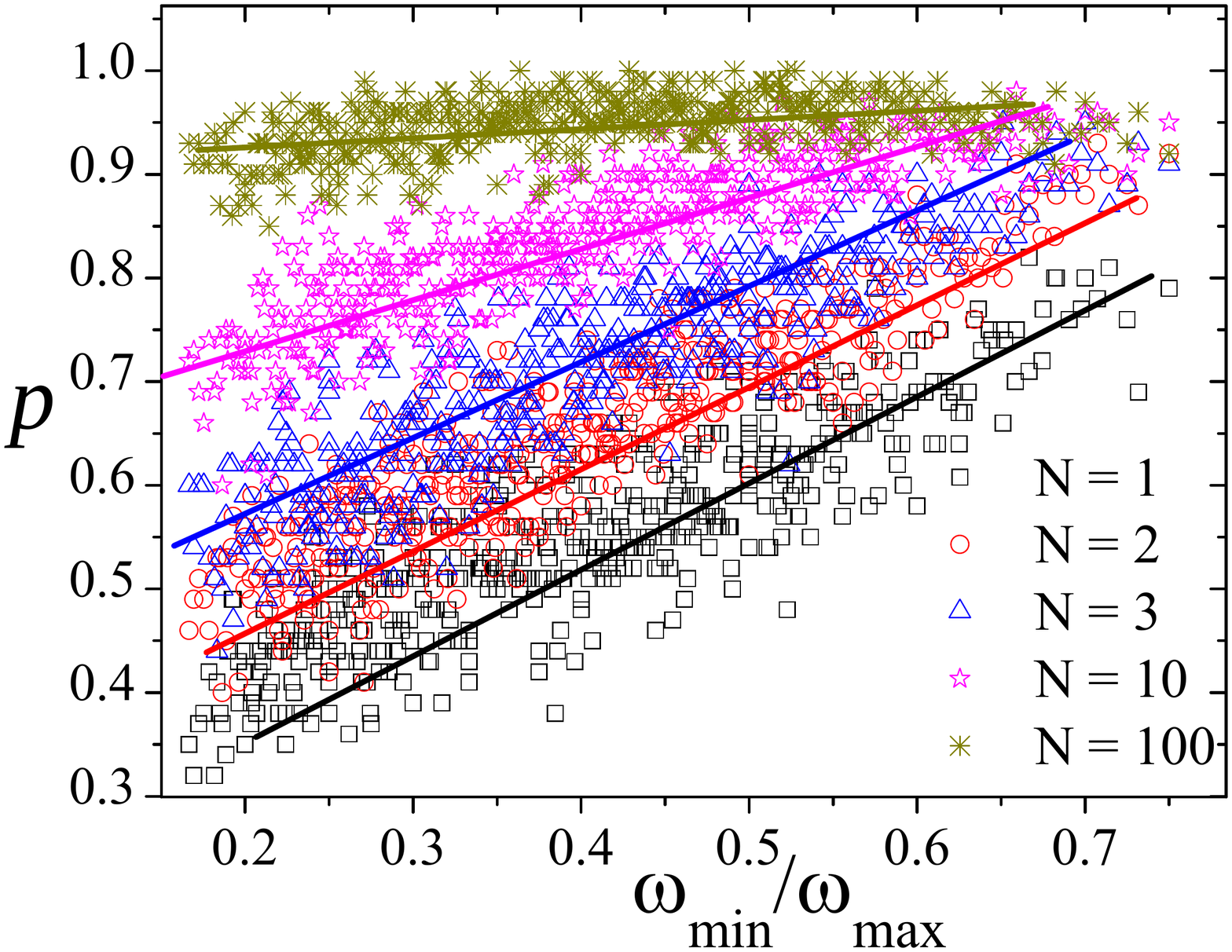}
\end{center}
\caption{Survival probability ($p$) as function of $\protect\omega _{\min }/%
\protect\omega _{\max }$ for different number of sines. We can observe that $%
p$ has a positive correlation with enlarge of $\protect\omega _{\min }$.
Here, we also used $l\protect\overset{N}{=}g=9.8$ m/s$^{2}$ and the results
were obtained via numerical integration. }
\label{Fig:number_of_cosines2}
\end{figure}

Figure \ref{Fig:Effects_number_cosine2} shows the survival probability for
different minimal/maximal frequency pairs used in this work. After some
numerical tests, we adopted $100\leq \omega _{\min }\leq 300$ while $400\leq
\omega _{\max }\leq 600$. In order to estimate the survival probability $p$,
we used $N_{run}=100$ repetitions. The plots differ in the number $N$ of
sines used. According to the adopted rainbow color scale, we can observe
that as $N$ enlarges, we have larger $p$. In all the simulations we used $l%
\overset{N}{=}g=9.8$ $m/s^{2}$.

A plot of $p$ as function of the ratio $\omega _{\min }/\omega _{\max }$ is
shown in Fig. \ref{Fig:number_of_cosines2}. We can observe that $p$ has a
positive correlation showing that the larger the minimum frequency the more
stability is expected.

Following, we test if it is possible to give more flexibility to the
stability criteria based previously on fixing the amplitude, according to
Eq. (\ref{Eq:Critical_A_c}). In order to proceed this analysis, we calculate
the ratio $\left\langle \tau \right\rangle /\tau _{\max }$ considering $%
A=A_{c}-\varepsilon $ and look into the deviations from stability as the
deviation parameter $\varepsilon $ gets larger. Here, $\left\langle \tau
\right\rangle $ is survival time averaged by $N_{run}$ repetitions. For
that, we consider $\varepsilon =0,2$ and $10$. In order to depict this
behavior, a plot of $\left\langle \tau \right\rangle /\tau _{\max }$ as
function of the ratio $\omega _{\min }/\omega _{\max }$ is shown in Fig. \ref%
{Fig:epsilon_effects}.

\begin{figure}[th]
\begin{center}
\includegraphics[width=1.0\columnwidth]{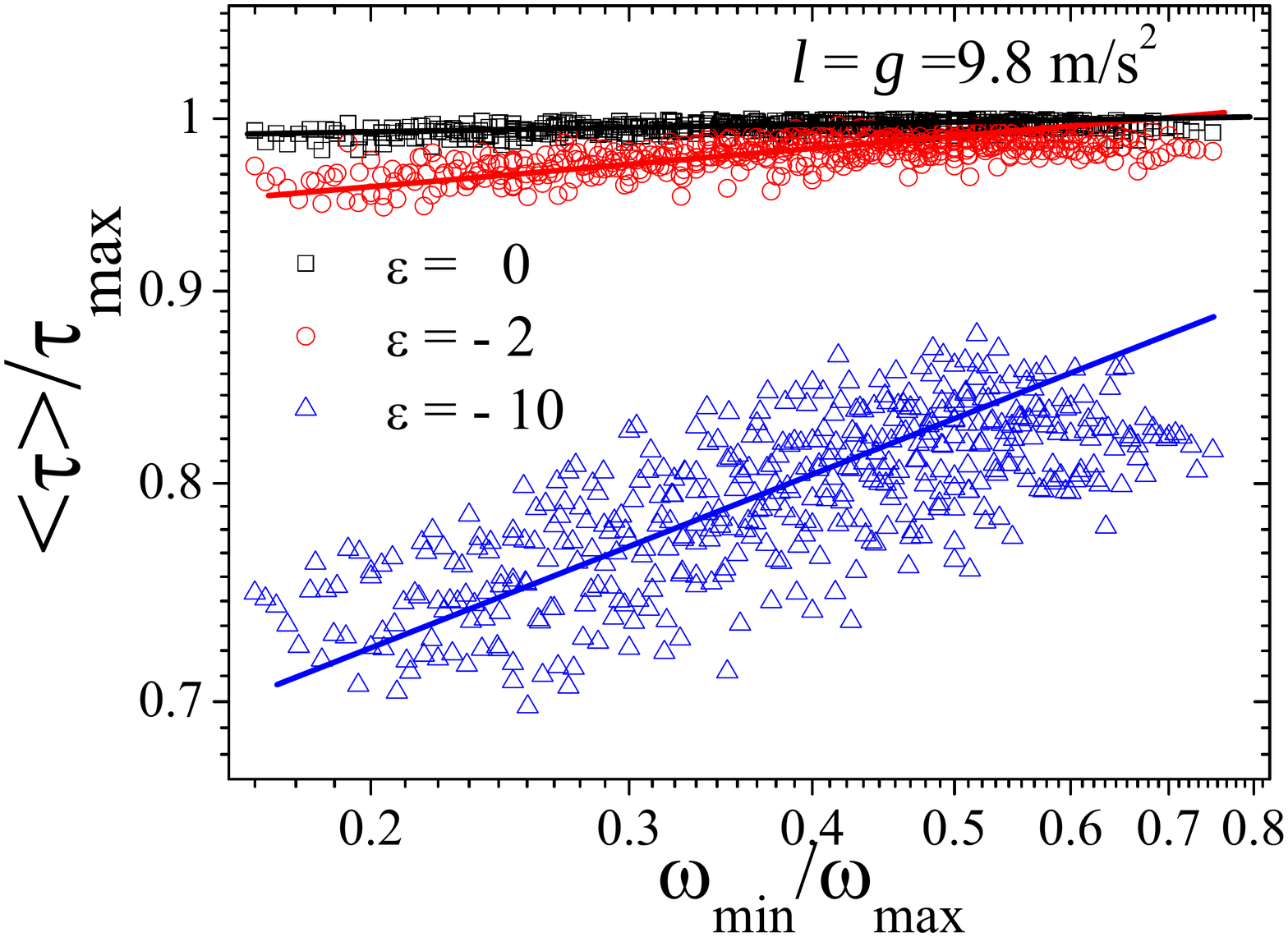}
\end{center}
\caption{Study of a more flexible stability criteria for the amplitude
according to Eq. (\protect\ref{Eq:Critical_A_c}). We can observe that the
average of survival time approximates of $\protect\tau _{\max }$ as $\protect%
\varepsilon \rightarrow 0$. Here, we also used $l\protect\overset{N}{=}g=9.8$
m/s$^{2}$ and $N=100$ and the results were obtained via numerical
integration.}
\label{Fig:epsilon_effects}
\end{figure}
It is important to mention that we can observe a positive correlation in the
increase of $\left\langle \tau \right\rangle $ when the minimal frequency
increases, corroborating the idea that the larger the minimum frequency the
more stability is expected.

So far, we have not explored the effects of pendulum size. So, we elaborated
similar diagrams presented in Fig. \ref{Fig:Effects_number_cosine2}, but now
changing the pendulum size $l$ in order to study how the stability of the
pendulum depends on this parameter.

\begin{figure*}[th]
\begin{center}
\includegraphics[width=0.5\columnwidth]{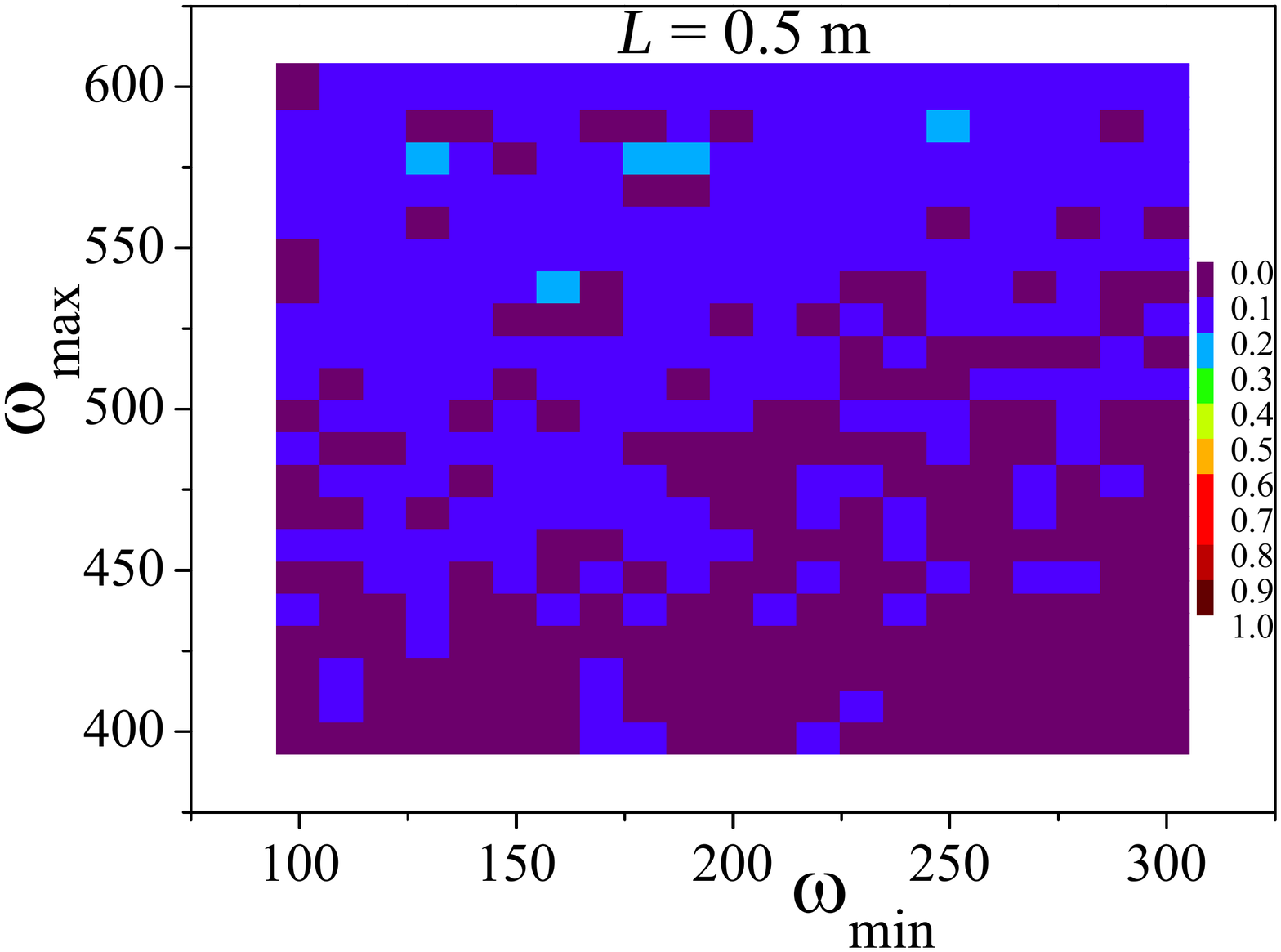}\includegraphics[width=0.5%
\columnwidth]{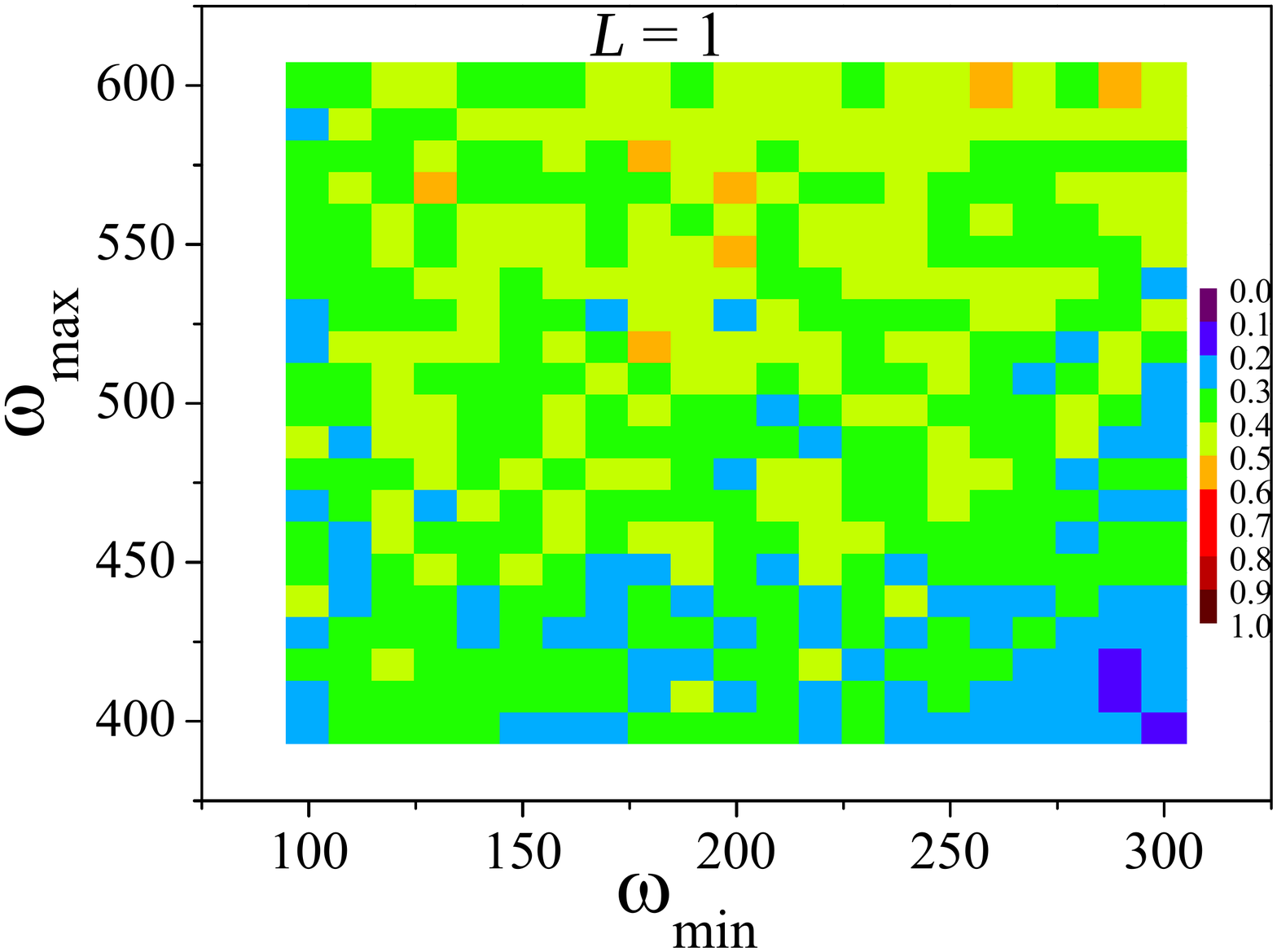} \includegraphics[width=0.5\columnwidth]{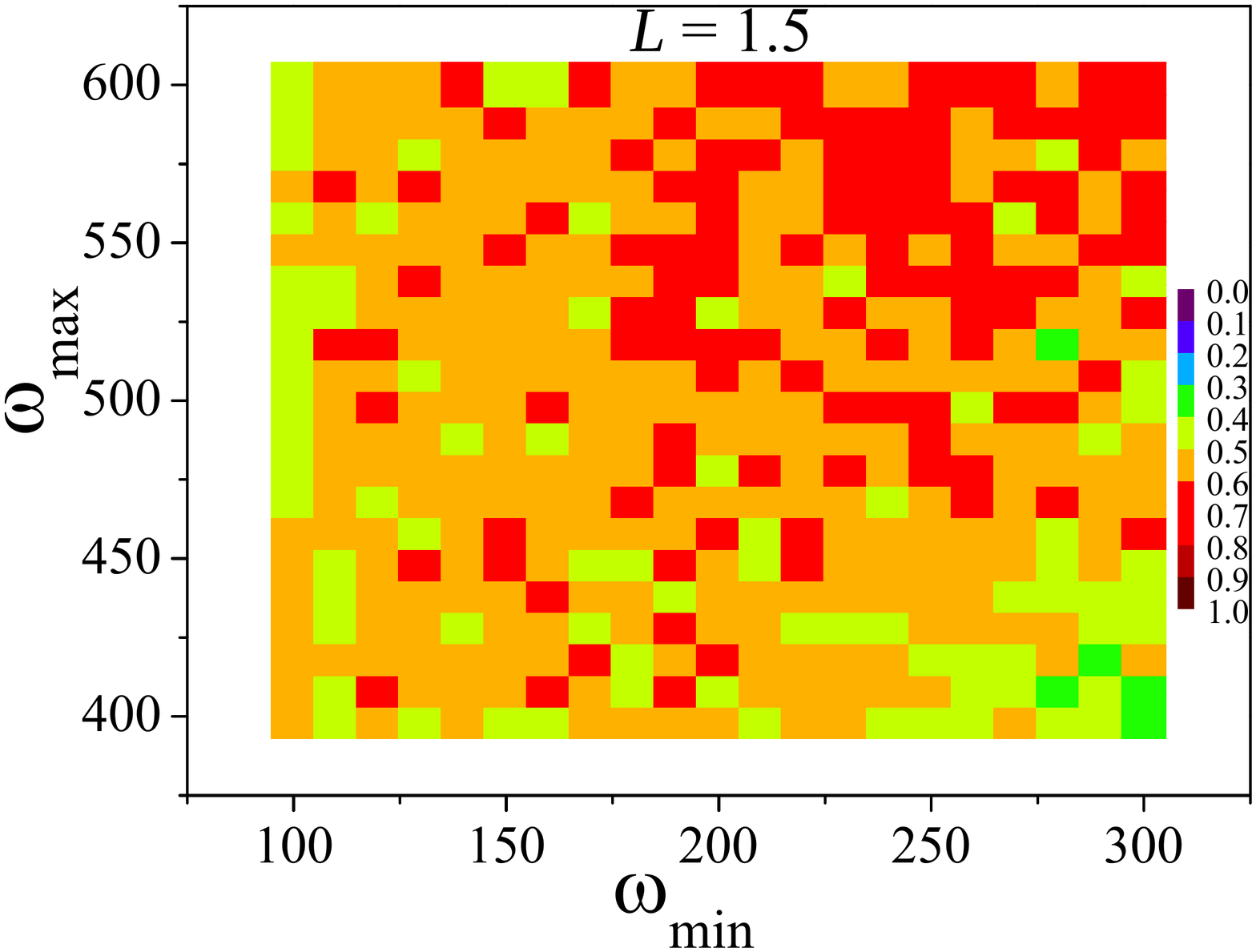}%
\includegraphics[width=0.5\columnwidth]{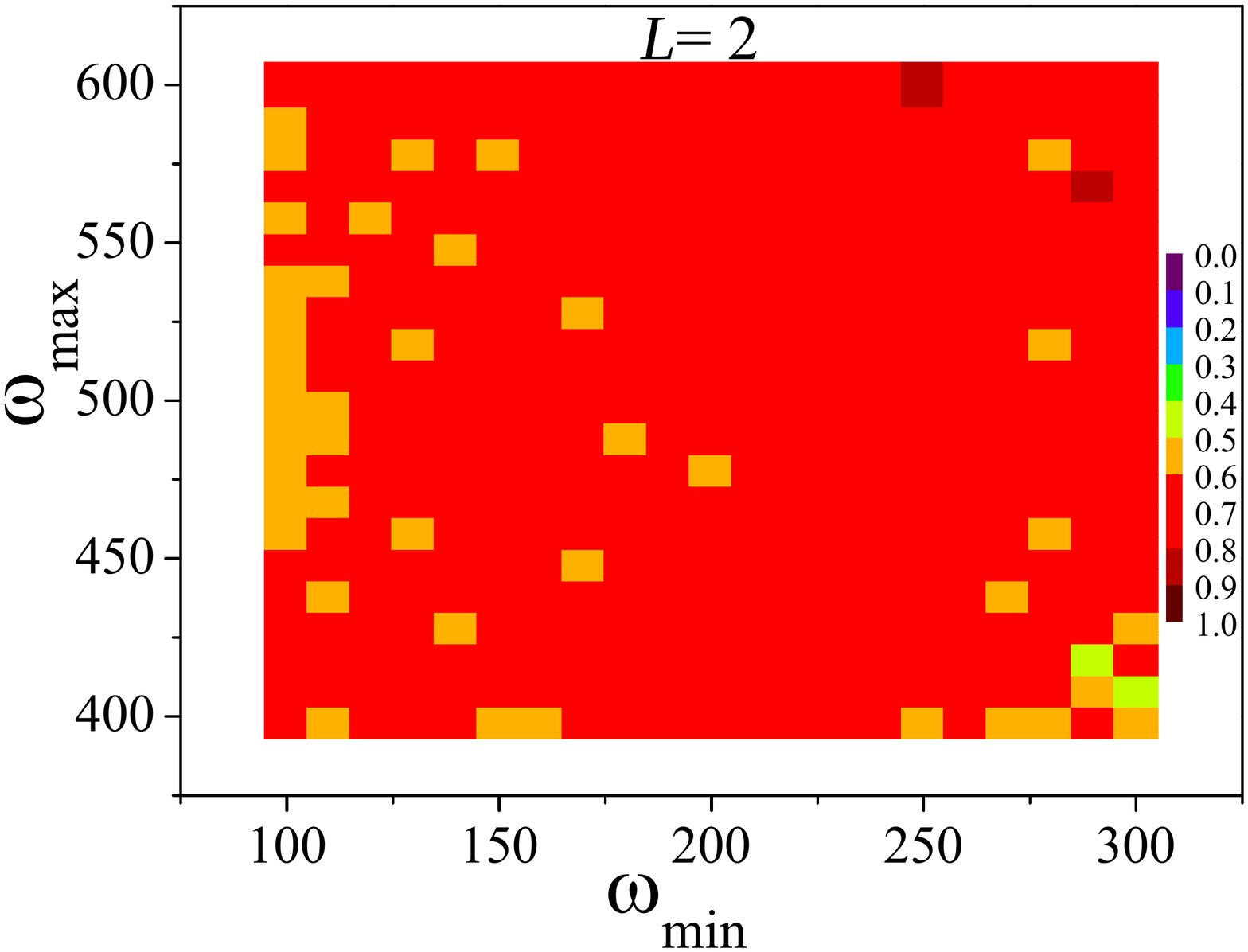}
\end{center}
\caption{Stability diagrams for different $l$ sizes obtained for numerical
integration and keeping $N=100$. For each pair ($\protect\omega _{\min },%
\protect\omega _{\max }$) we draw the survival probability $p$ according to
rainbow color scale (very cold $p\rightarrow 0$ and very hot $p\rightarrow 1$%
). }
\label{Fig:size_effects}
\end{figure*}

We can observe that larger pendulums are most likely stabilized. Table \ref%
{Table:I} shows the maximum survival probability obtained for all pairs $%
\left( \omega _{\min },\omega _{\max }\right) $ used in the plots of Fig. %
\ref{Fig:size_effects} for each pendulum size ($l$) studied. In columns 2
and 3 we show the values of $\omega _{\min }$ and $\omega _{\max }$ for
which $p_{\max }$ was found. We can observe a systematic increase of $%
p_{\max }$ as $l$ increases. For $l=9.81$ m, for example, we obtain $p_{\max
}=1$.

\begin{table}[tbp] \centering%
\begin{tabular}{llll}
\hline
$l$ (m) & $\omega _{\min }$ & $\omega _{\max }$ & $p_{\max }$ \\ \hline
0.5 & 180\qquad & 580 & 0.24 \\ 
1.0 & 200 & 550 & 0.34 \\ 
1.5 & 190 & 540 & 0.50 \\ 
2.0 & 250 & 600 & 0.64 \\ 
2.5 & 250 & 600 & 0.72 \\ 
\textbf{9.8} & \textbf{230} & \textbf{440} & \textbf{1.00} \\ \hline
\end{tabular}%
\caption{Maximum survival probability for different lengths $l$ of the 
pendulum}\label{Table:I}%
\end{table}%

So, we can conclude that the inequality given by Eq. (\ref{Eq:Critical_A_c})
is suitable and therefore can be used in the study of inverted pendulums.
This conclusion is supported by numerical simulations which, in addition,
describe the effects of $N$ and the size of pendulum on its stability. It is
important to notice that deviations of this inequality can be observed for
small pendulums but its validity occurs for larger $l$-values.

The next subsection deals with the validity of Eq. (\ref{Eq:Critical_A_c})
for a particular set of parameters and show the range in which the critical
amplitude $A_c$ is valid.

\subsection{Exploring the range of validity of the critical amplitude and
scaling}

\begin{figure*}[th]
\begin{center}
\includegraphics[width=0.52\columnwidth]{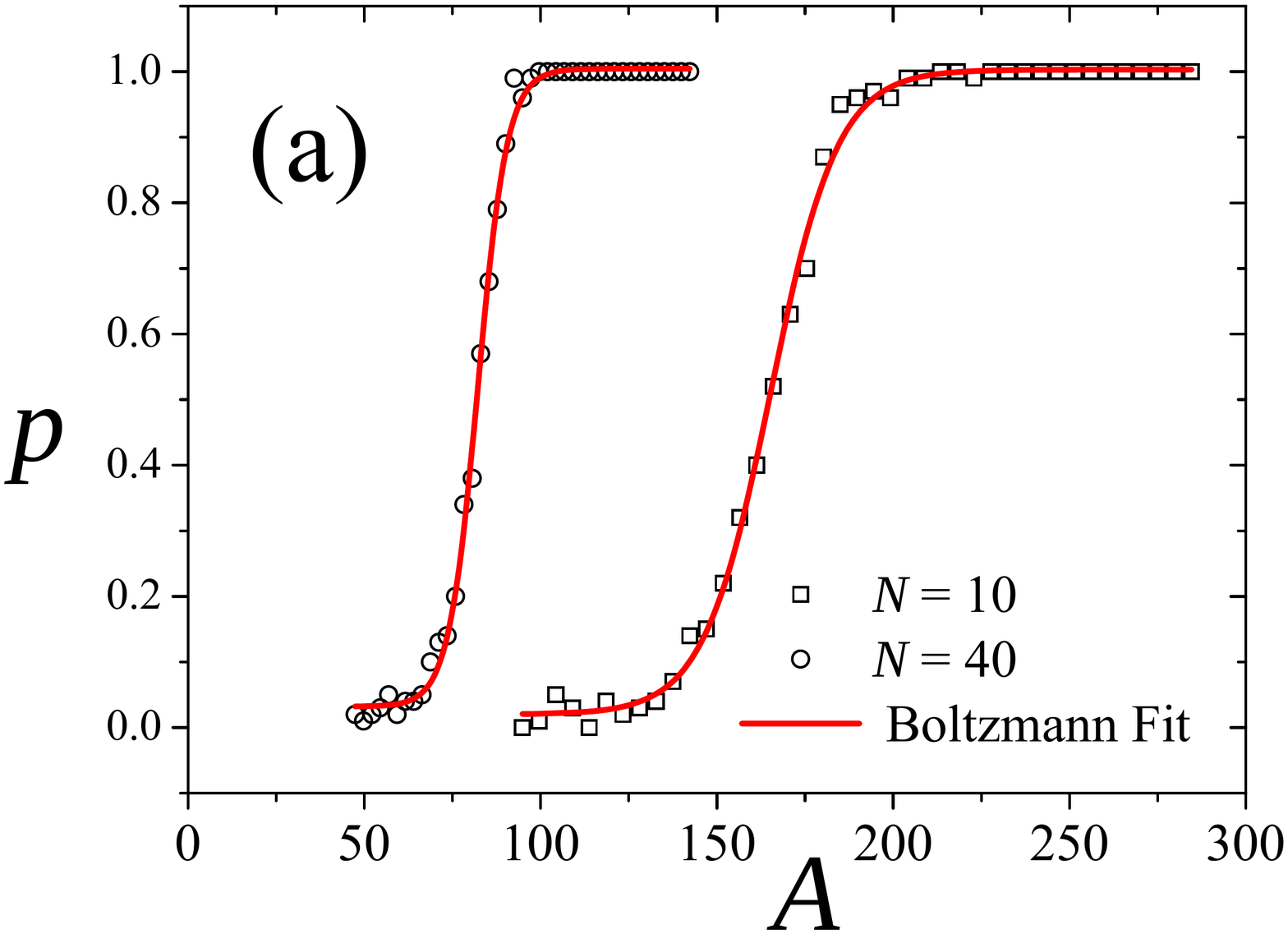}%
\includegraphics[width=0.52\columnwidth]{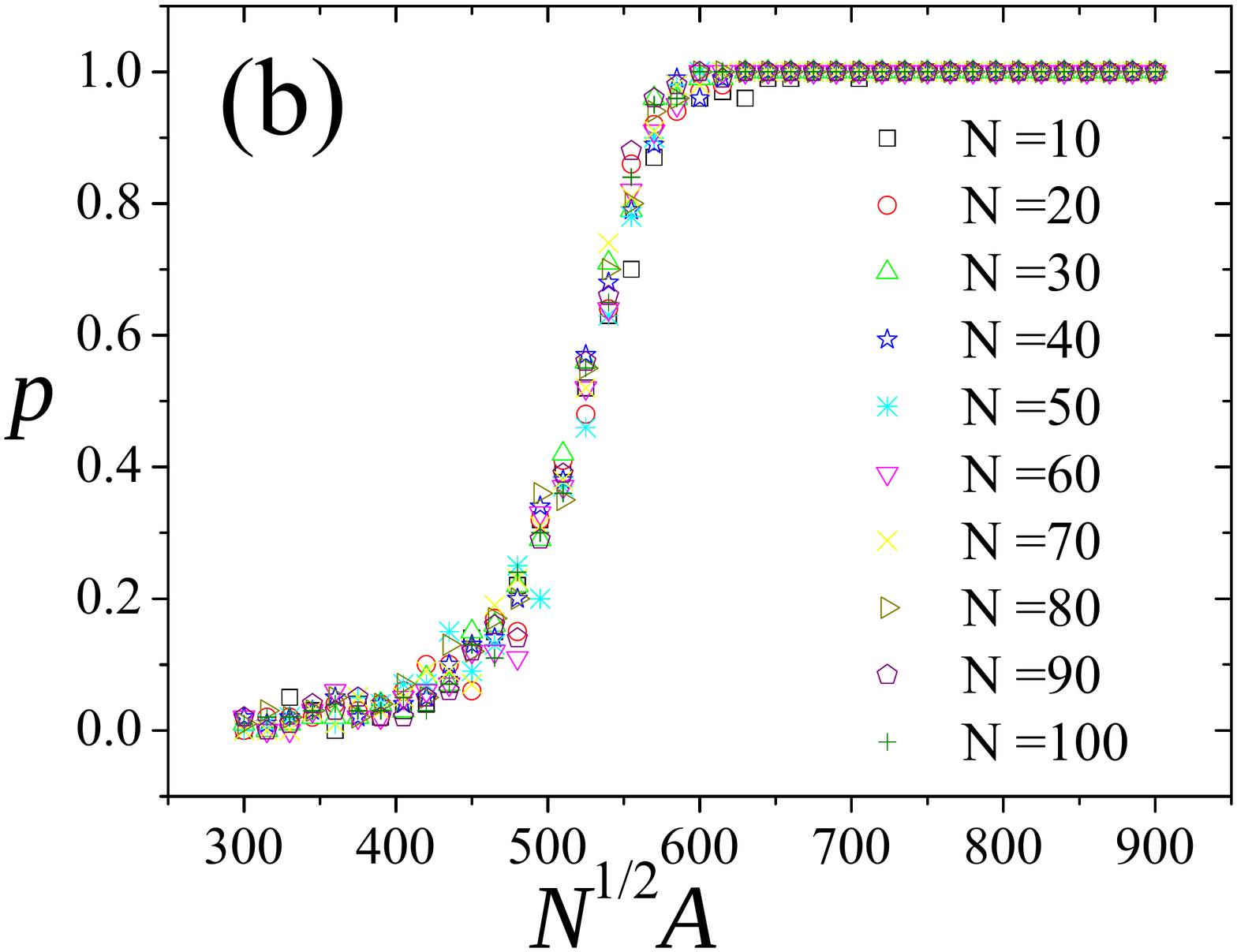}
\end{center}
\caption{(a) Survival probability $p$ as function of the amplitude $A$ for
two different number of sines used in numerical results: $N=10$ and $N=40$.
The points correspond to numerical results while the lines correspond to the
Boltzmann fits. (b) Collapse of curves $p\times N^{1/2}A$. We can observe
that all curves, corresponding to different number of sines, become
practically the same universal curve, under scaling $A\rightarrow N^{1/2}A$.}
\label{Fig:Fit_and_Scaling}
\end{figure*}

Let us now explore in more detail the validity of the critical amplitude,
previously calculated via perturbative analysis, by considering numerical
simulations. This study makes possible to better explore the scaling
properties of the survival probability ($p$). To perform the simulations, we
fixed the frequencies in $\omega _{\min }=300$ and $\omega _{\max }=600$ and
set $l\overset{N}{=}g=9.8$ $m/s^{2}$, for which we are sure that $p_{\max }=1
$ (see Table \ref{Table:I}). We also used the same number of repetitions (as
in the previous analysis) to estimate $p$: $N_{run}=100.$ We performed
numerical integrations considering several different number of sines, $10
\leq N \leq 100$, with $\Delta N=10$, in order to obtain $p$ as function of
the amplitude $A$. First, we show in Fig. \ref{Fig:Fit_and_Scaling} (a) two
curves for $N=10$ and $N=40$ as function of $A$. 
The typical transition from $p=0$ to $p=1$ suggests a familiar fit, known as
Boltzmann curve which is parameterized as 
\begin{equation}
p=p_{\min }+\frac{(p_{\max }-p_{\min })}{1+\exp \left[ \frac{(A-A_{0})}{%
\Delta }\right]. }  \label{Eq:Boltzmann}
\end{equation}
Therefore we performed fits according to this function as shown in Fig. \ref%
{Fig:Fit_and_Scaling} (a) by the lines in red which can be compared with the
points obtained from our numerical integration. The nonlinear fit, taking
into account the Levenberg--Marquardt algorithm (see for example \cite%
{Press1992}), yields for $N=10$, the parameters $p_{\min }=0.020(7)$, $%
p_{\max }=1.003(5)$, $A_{0}=165.3(4)$, and $\Delta =9.5(3)$, while for $N=40$%
, we obtained the parameters $p_{\min }=0.032(6)$, $p_{\max}=1.004(4)$, $%
A_{0}=82.21(17)$, and $\Delta =4.15(15)$. We also obtained an excelent fit
with the coefficient of determination $\alpha =0.998$ in both cases (the
closer to 1, the better is the fit).

Another important question here is that the analytical result from Eq. \ref%
{Eq:Critical_A_c} indicate that $A_{c}\sim N^{-1/2}$. This suggests a
scaling relation for the survival probability: 
\begin{equation*}
p(A,N)=h(b^{-1/2}A,bN).
\end{equation*}
By imposing the scaling $bN=1$, we have$p(A,N)=h(N^{1/2}A,1)$ where $%
h(x,1)=h(x)$ has the property: 
\begin{equation*}
h(x)=\left\{ 
\begin{array}{ccc}
1 & \text{if} & x\rightarrow \infty \\ 
&  &  \\ 
0 & \text{if} & x\rightarrow 0%
\end{array}%
\right.
\end{equation*}

\begin{figure*}[th]
\begin{center}
\includegraphics[width=1.0\columnwidth]{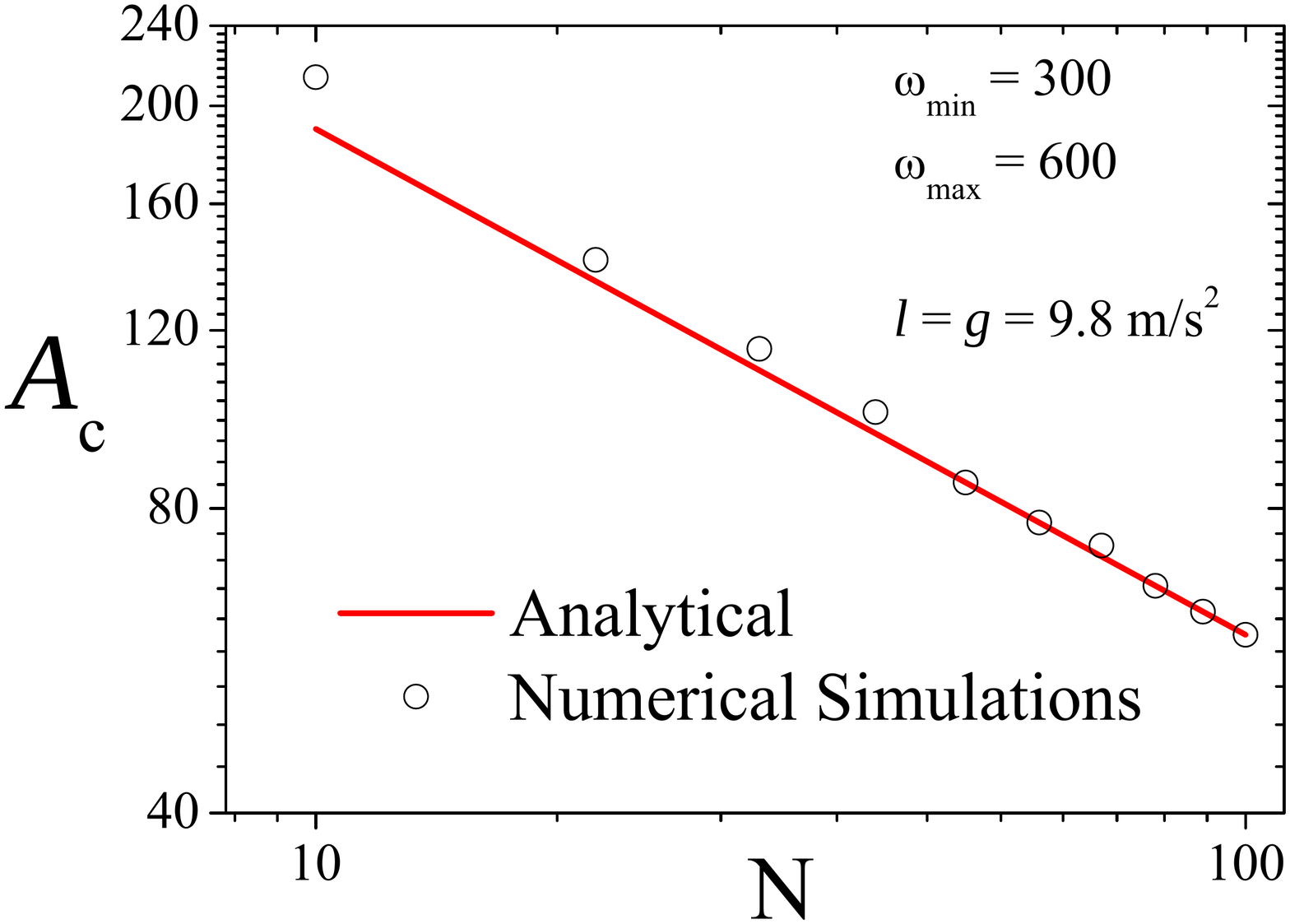}
\end{center}
\caption{Log-log plot of $A_{c}\times N$. The points correspond to numerical
estimates while the line corresponds to analytical estimates. We can see a
good agreement between the results which becomes better as $N$ increases.}
\label{Fig:Ac_versus_N}
\end{figure*}

In Fig. \ref{Fig:Fit_and_Scaling} (b) we show the curves $p\times N^{1/2}A$
for different values of $N$. We can observe a collapse of all curves through
the finite size scaling considered according to our previous description.
But what to say about $A_{c}$? Can we obtain the analytical prediction
described by Eq. \ref{Eq:Critical_A_c} from these numerical results? Yes.
Actually the procedure is very simple. We can numerically estimate $A_{c}$
from the plots $p\times A$ by taking the first value of $A$ (the numerical
critical amplitude) such that the survival probability is exactly equal to
1. After, we can compare this numerical result with $A_{c}$ obtained
analitically through Eq. \ref{Eq:Critical_A_c}. Figure \ref{Fig:Ac_versus_N}
shows a good agreement between these two analysis which, as can be seen,
becomes better for larger values of $N$.

Thus, our results corroborate the analytical estimate for the critical
amplitude obtained from perturbation theory. In addition, the scaling $%
A\rightarrow N^{1/2}A$ produces a collapse of all survival probabilities.
The same procedure can be repeated for smaller pendulums. In this case $%
p_{\max}$ may be less than 1 and the numerical procedures for determining $%
A_{c}$ must be more careful.

\section{Summary and brief conclusions}

\label{Section:Conclusions}

By showing an interesting extension of results obtained by Dettman \textit{%
et al.} \cite{Prado2004} in the context of the stabilization of cosmological
photons which are far from parameters of a real pendulum, we fixed $g=9.8$ $%
m/s^{2}$ and showed that an inverted pendulum can
``stochastically\textquotedblright\ be stabilized under of superposition of
sines if the amplitudes are rescaled according to square frequencies.

In our analytical study, we were able to obtain the critical amplitude which
depends on the maximum and minimum amplitudes used as parameters to
uniformly sort frequencies which were numerically tested considering
different number of sines. Our numerical results corroborate the critical
lower bound amplitude obtained analytically and, in addition, bring
important details about its applicability which cannot be captured by the
perturbative analysis. The results show, for example, that as pendulum size
increases, more prominent is the verification of the analytical result to
the amplitude. We also verify that the number of sines directly impacts on
the verification of the lower bound. Moreover, deviations from these
theoretical predictions have direct effects in numerical simulations which
are observed by tuning a deviation parameter $\varepsilon $ introduced in
our analysis. We also conducted an interesting study about finite-size
scaling observing the survival probability $p$ as function of the amplitudes 
$A$. First, we showed that $p \times A$ follows a Boltzmann function moving
from $p=0$ ($A<A_{c}$) to $p=1$ ($A>A_{c}$). In addition, our results show
that for a large values of $l$, the dependence of $A_{c}$ on $N$, predicted
by perturbative analysis, fits very well the numerical results.

It is interesting to say that our analysis does not depend on $\alpha $ that
appears in Eq. (\ref{Eq:General_solution}) showing that other dependences
can be tested further. It is also important to mention that other
distributions of frequencies $\omega $ may be better explored in the future.

\textbf{Acknowledgments --} This research work was in part supported
financially by CNPq (National Council for Scientific and Technological
Development). R. da Silva would like to thank Prof. L.G. Brunnet (IF-UFRGS)
for kindly providing the computational resources from Clustered Computing
(cluster-slurm.if.ufrgs.br))

\end{document}